\documentclass[useAMS,usenatbib,onecolumn]{mn2e}

\usepackage{dcolumn}
\usepackage{graphicx}
\usepackage{url}
\usepackage{color}
\usepackage{epstopdf}
\usepackage{textcomp}
\usepackage[T1]{fontenc}

\newcommand{\cm}{cm$^{-1}$}

\newcommand{\ai}{\textit{ab initio}}

\newcommand{\eqref}[1]{(\ref{#1})}

\newcommand{\p}{^\prime}
\newcommand{\pp}{^{\prime\prime}}

\title[ExoMol molecular line lists V: NaCl and KCl]{ExoMol molecular line lists V: The ro-vibrational spectra of NaCl and KCl}

\date{\today}
\author[Barton et al]{Emma J. Barton$^{1}$, Christopher Chiu$^{1}$, Shirin Golpayegani$^{1}$, Sergei N. Yurchenko$^{1}$,
\newauthor  Jonathan Tennyson$^{1}$, Daniel J. Frohman$^{2}$ and Peter F. Bernath$^{2}$ \\
$^{1}$Department of Physics and Astronomy, University College London, London WC1E 6BT, UK; \\
$^{2}$Department of Chemistry and Biochemistry, Old Dominion University, Norfolk 23529-0126, USA}
\date{Accepted XXXX. Received XXXX; in original form XXXX}

\pagerange{\pageref{firstpage}--\pageref{lastpage}} \pubyear{2014}

\begin{document}

\maketitle

\begin{abstract}

  Accurate rotation-vibration line lists for two molecules, NaCl and
  KCl, in their ground electronic states are presented. These line
  lists are suitable for temperatures relevant to exoplanetary
  atmospheres and cool stars (up to 3000 K). Isotopologues
  $^{23}$Na$^{35}$Cl, $^{23}$Na$^{37}$Cl, $^{39}$K$^{35}$Cl, $^{39}$K$^{37}$Cl,
  $^{41}$K$^{35}$Cl and $^{41}$K$^{37}$Cl are considered.
  Laboratory data was used to refine \ai\ potential energy curves in
  order to compute accurate ro-vibrational energy levels. Einstein A
  coefficients are generated using newly determined \ai\ dipole moment
  curves calculated using the CCSD(T) method. New
  Dunham Y$_{ij}$ constants for KCl are generated by a reanalysis of
  a  published Fourier transform infrared emission spectra. Partition
  functions plus full line lists of ro-vibration transitions are made
  available in an electronic form as supplementary data to this
  article and at \url{www.exomol.com}.

\end{abstract}

\textit{molecular data; opacity; astronomical data bases: miscellaneous; planets and satellites: atmospheres; stars: low-mass}

\label{firstpage}

\section{Introduction}

NaCl and KCl are important astrophysical species
as they are simple, stable molecules containing atoms of
relatively high cosmic abundance. Na, K and Cl are the 15th, 20th and
19th most abundant elements in the interstellar medium \citep{04CaLeMu.KCl}. In fact NaCl
could be as abundant as the widely-observed SiO molecule \citep{87CeGuxx.both}.
\citet{87CeGuxx.both} reported the first detection of metal halides,
NaCl, KCl, AlCl and, more tentatively, AlF, in the circumstellar
envelope of carbon star IRC+10216. These observations have been
followed up recently by \citet{12AgFoCe.both}, who also observed CS,
SiO, SiS and NaCN. NaCl has also been detected in the circumstellar
envelopes of oxygen stars IK Tauri and VY Caris Majoris
\citep{07MiApWo.NaCl}. Another environment in which these molecules
have been found is the tenuous atmosphere of Jupiter's moon Io.
Submillimetre lines of NaCl, and more tentatively KCl, were detected by
\citet{03LePaMo.NaCl} and \citet{13MoLeMo.KCl} respectively. NaCl has
also been identified in the cryovolcanic plumes of Saturn's moon
Enceladus alongside its constituents Na and Cl \citep{11PoScHi.NaCl}.
K was also detected but the presence of KCl could not be confirmed.
Furthermore NaCl and KCl are expected to be present in Super Earth
atmospheres \citep{12ScLoFe.both} and may form in the observable
atmosphere of the known object GJ1214b \citep{14KrBeDe.both}.

The alkali chlorides are also of industrial importance as they are
products of coal and straw combustion. Their presence in coal
increases the rate of corrosion in coal fired power plants
\citep{14YaKaLi.KCl}.  Therefore it is important to monitor their
concentrations, which can be done spectroscopically provided
the appropriate data is available.

The importance of NaCl and KCl spectra have motivated a number of laboratory
studies, for example \citet{57RiKlxx.NaCl, 54HoMaSt.KCl,
  88HoFuNa.NaCl, 89UeHoNa.NaCl, 64ClGoxx.both, 90UeHoKo.KCl}. The most
recent and extensive research on KCl and NaCl spectra has been
performed by \citet{97RaDuGu.NaCl}, whom investigated infrared
emission lines of Na$^{35}$Cl, Na$^{37}$Cl and $^{39}$K$^{35}$Cl,
\cite{04CaLeMu.KCl}, whom measured microwave and millimetre wave lines
of $^{39}$K$^{35}$Cl, $^{39}$K$^{37}$Cl, $^{41}$K$^{35}$Cl,
$^{41}$K$^{37}$Cl and $^{40}$K$^{35}$Cl, and \cite{02CaLeWi.NaCl},
whom recorded microwave and millimetre wave lines of Na$^{35}$Cl and
Na$^{37}$Cl.

Dipole moment measurements have been carried out by
\citet{70LeWaDy.NaCl} for Na$^{35}$Cl and Na$^{37}$Cl,
\citet{67WaDyxx.KCl} for $^{39}$K$^{35}$Cl and $^{39}$K$^{37}$Cl, and
\citet{68HeLoMe.both} for $^{39}$K$^{35}$Cl, Na$^{35}$Cl and Na$^{37}$Cl.

It appears the only theoretical transition line lists for these
molecules are catalogued in the Cologne Database for Molecular
Spectroscopy (CDMS), see
\citet{cdms}. They were constructed using data reported in
\citet{02CaLeWi.NaCl, 64ClGoxx.both, 89UeHoNa.NaCl, 70LeWaDy.NaCl} for
NaCl, and \citet{04CaLeMu.KCl, 64ClGoxx.both, 67WaDyxx.KCl} for KCl.
The lists are limited to $v$ = 4, $J$ = 159 and do not include a list
for $^{41}$K$^{37}$Cl. In this paper we aim to compute more
comprehensive line lists for the previously studied isotopologues and
the first theoretical line list for $^{41}$K$^{37}$Cl.

The ExoMol project aims to provide line lists on all the molecular
transitions of importance in the atmospheres of planets. The aims,
scope and methodology of the project have been summarised by
\citet{jt528}. Lines lists for X~$^{2}\Sigma^{+}$ XH molecules, X = Be,
Mg, Ca, and X~$^{1}\Sigma^{+}$ SiO have already been published
(\citet{jt529, jt563} respectively). In this paper, we present
ro-vibrational transition lists and associated spectra for two NaCl
and four KCl isotopologues.

\section{Method}

The nuclear motion Schr\"{o}dinger equation allowing for
Born-Oppenheimer Breakdown (BOB) affects, is solved for species XCl
using program LEVEL \citep{lr07}. To initiate these calculations,
program DPOTFIT \citep{dpotfit} was used to generate a refined
potential energy curve (PEC) for each molecule by fitting \ai\ curves
to laboratory data.

\subsection{Spectroscopic Data}

The most comprehensive and accurate sets of available laboratory
measurements are the infrared ro-vibrational emission lines of
\citet{97RaDuGu.NaCl} and the microwave rotational lines of
\citet{02CaLeWi.NaCl} and \citet{04CaLeMu.KCl} all of which were
recorded at temperatures in the region of 1000 C, see
Table~\ref{tab:obsdata}. For KCl Fourier transform infrared emission
spectra measured by \citet{97RaDuGu.NaCl} has been re-analysed and
re-assigned as part of this work, see Section 2.2. The Dunham
constants (Y$_{ij}$) obtained from this new fit are provided in
Table~\ref{tab:dunham}.  Ro-vibrational emission lines derived from
this were used in place of those presented by \citet{97RaDuGu.NaCl}
because of problems found in the previous analysis.

\subsection{Reanalysis of the KCl infrared spectrum}

\citet{97RaDuGu.NaCl} reported spectroscopic constants derived from an
infrared emission spectrum of KCl recorded with a high resolution
Fourier transform spectrometer (FTS). By using the new constants derived
from the millimetre wave spectrum by \citet{04CaLeMu.KCl} to simulate
the infrared spectrum of $^{39}$K$^{35}$Cl with PGOPHER
\citep{PGOPHER}, it was clear that \citet{97RaDuGu.NaCl} had
mis-assigned much of the complex spectrum. The \citet{97RaDuGu.NaCl}
spectrum was therefore re-analyzed.  As a first step, the millimetre
wave line list from \citet{04CaLeMu.KCl} was refitted with the
addition of two Dunham parameters, Y$_{23}$ and Y$_{41}$.  These
parameters were found to improve the quality of the fit. The
\citet{04CaLeMu.KCl} constants plus Y$_{23}$ and Y$_{41}$ were then
used to calculate band constants used as input for PGOPHER.  Using
PGOPHER the infrared line positions were selected manually and then
refitted along with the Caris data using our LSQ fit program. There
were 253 R branch lines of $^{39}$K$^{35}$Cl fit from the 6-5, 5-4,
4-3, 3-2, 2-1, and 1-0 bands, and the Y$_{10}$, Y$_{20}$ and Y$_{30}$
vibrational constants were added.  The quality of the observed
spectrum was insufficient to fit additional bands or P branch lines.
The final constants from our global fit are compared to the values
reported by \citet{04CaLeMu.KCl} in Table~\ref{tab:dunham}. The
Y$_{10}$ and Y$_{20}$ ($\omega_{e}$ and $-\omega_{e}x_{e}$) constants
of \citet{04CaLeMu.KCl} were derived entirely from millimetre wave
data using Dunham relationships and are in good agreement with the
values we have determined directly from infrared observations.

\begin{table}
\caption{Summary of laboratory data used to refine the KCl and NaCl
potential energy curves. Temperatures are those given in the cited papers.
Uncertainties are the maximum quoted uncertainty given in the cited papers.}
\label{tab:obsdata} \footnotesize
\begin{center}
\begin{tabular}{llccl}
\hline
Reference&    Transitions&   Frequency range    & Uncertainty \\
         &               &  (cm$^{-1}$)        & (cm$^{-1}$) \\
\hline
\citet{02CaLeWi.NaCl}    & $\Delta v=0$, $\Delta J = \pm 1$&  6.6 -- 31  & $6.7 \times 10^{-6}$\\
       & Na$^{35}$Cl, $v = 0 - 5$, $J \leq 72$ &&& \\
       & Na$^{37}$Cl, $v = 0 - 4$, $J \leq 76$ &&& \\
       \\
\citet{04CaLeMu.KCl}     & $\Delta v=0$, $\Delta J = \pm 1$&  5.6 -- 31  & $6.7 \times 10^{-6}$\\
       & $^{39}$K$^{35}$Cl, $v = 0 - 7$, $J \leq 127$ &&& \\
       & $^{39}$K$^{37}$Cl, $v = 0 - 7$, $J \leq 129$ &&& \\
       & $^{41}$K$^{35}$Cl, $v = 0 - 6$, $J \leq 128$ &&& \\
       & $^{41}$K$^{47}$Cl, $v = 0 - 5$, $J \leq 131$ &&& \\
       \\
\citet{97RaDuGu.NaCl} & $\Delta v=1$, $\Delta J = \pm 1$&  240 -- 390  & 0.005\\
       & Na$^{35}$Cl, $v = 0 - 8$, $J \leq 118$ &&& \\
       & Na$^{39}$Cl, $v = 0 - 3$, $J \leq 91$ &&& \\
       \\
This work & $\Delta v=1$, $\Delta J = + 1$&  240 -- 390  & 0.005\\
       & $^{39}$K$^{35}$Cl, $v = 0 - 6$, $J \leq 131$ &&& \\
\hline
\end{tabular}
\end{center}

\end{table}

\begin{table}
\caption{Dunham Constants (in \cm) of the X~$^{1}\Sigma^{+}$ state of KCl.
(Uncertainties are given in paranthesis in units of the last digit.)}
\label{tab:dunham} \footnotesize
\begin{center}
\begin{tabular}{lrr}
\hline
Constant & This Work & \citet{04CaLeMu.KCl} \\
\hline
Y$_{01}$	&	0.1286345842(27)	&	0.1286345835(38)	\\
Y$_{11}$	&	-7.896827(31)E-4	&	-7.896870(24)E-4	\\
Y$_{21}$	&	1.5916(14)E-6	&	1.59637(59)E-6	\\
Y$_{31}$	&	5.47(27)E-9	&	4.297(50)E-9	\\
Y$_{41}$	&	-7.6(18)E-11	&	-	\\
Y$_{02}$	&	-1.0868336(42)E-7	&	-1.0868276(72)E-7	\\
Y$_{12}$	&	-1.112(19)E-11	&	-1.184(13)E-11	\\
Y$_{22}$	&	3.729(30)E-12	&	3.851(15)E-12	\\
Y$_{03}$	&	-2.0955(35)E-14	&	-2.0975(64)E-14	\\
Y$_{13}$	&	3.614(65)E-16	&	3.877(37)E-16	\\
Y$_{23}$	&	4.4(10)E-18	&	-	\\
Y$_{04}$	&	-4.019(99)E-20	&	-4.04(19)E-20	\\
Y$_{10}$	&	279.88193(76)	&	279.889346(936)	\\
Y$_{20}$	&	-1.19671(26)	&	-1.1972502(793)	\\
Y$_{30}$	&	0.003094(26)	&	-	\\
\hline
\end{tabular}
\end{center}
\end{table}

\subsection{Dipole Moments}

Experimental measurements of the permanent dipole as a function of
vibrational state have been performed by \citet{70LeWaDy.NaCl,
  67WaDyxx.KCl, 68HeLoMe.both} who considered NaCl, KCl and both
molecules respectively. Additionally \citet{01Plxxxx.both} calculated
dipole moments at equilibrium bond length as part of theoretical study
comparing various levels of theory (SCF, MP2, CCSD and CCSD(T)). \citet{04GiYo}
computed NaCl dipole  moment curves (DMCs) using a multi-reference configuration
interaction (MRCI) approach and extrapolated basis sets.
However there appears to be no published KCl dipole moment curves (DMCs),
experimental or \ai.

\begin{figure}
\begin{center}
\scalebox{0.4}{\includegraphics{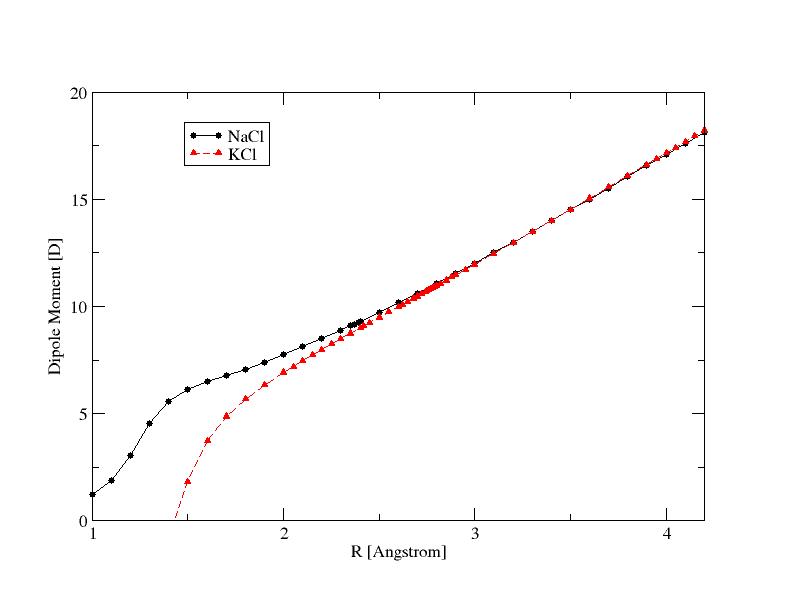}}
\caption{{\it Ab initio} dipole moment curves for NaCl and KCl in their ground electronic states.}
\label{fig:dipole}
\end{center}
\end{figure}

We determined new DMCs for both molecules using high-level \ai\
calculations, shown in Fig.~\ref{fig:dipole}. These were performed
using MOLPRO \citep{molpro.method}. The final NaCl DMC was computed
using an aug-cc-pCVQZ-DK basis set and the CCSD(T) method, where the
both core-valence and relativistic effects were also taken into
account. Inclusion of both effects is known to be important
\citep{jt573}. In case of KCl, an effective core potential ECP10MDF
(MCDHF+Breit) in conjunction with the corresponding basis set
\citep{05LiScMe.ai} was used for K and aug-cc-pV(Q+d)Z was used for
Cl.  In both cases the electric dipole moment were obtained using the
finite field method.  The \ai\ DMC grid points were used directly in
LEVEL. Eq0uilibrium bond length dipole moments are compared in
Table~\ref{tab:dipole}.  Our computed equilibrium dipole for KCl is
about 1~\%\ larger than the experimental value. For NaCl this
difference is closer to 2~\%\ but our final, CCSD(T) value is close to
those calculated by \citet{04GiYo}.

\begin{table}
\caption{Comparison of Na$^{35}$Cl and $^{39}$K$^{35}$Cl dipole moments at equilibrium internuclear distance.}
\label{tab:dipole}
\begin{center}
\begin{tabular}{llll}
\hline
Reference & Method & $\mu$ (NaCl) & $\mu$ (KCl)\\
          &        &   Debye    &  Debye   \\
\hline
\citet{68HeLoMe.both}    &  Experiment &  8.9721  &  10.2384  \\
\citet{01Plxxxx.both}    & SCF & 9.2774  & 10.6626  \\
\citet{01Plxxxx.both}    & MP2 & 9.0740  & 10.4923  \\
\citet{01Plxxxx.both}   & CCSD & 9.0715  & 10.4847  \\
\citet{01Plxxxx.both}   & CCSD(T) & 9.0257  & 10.4542  \\
This work   & CCSD(T) & 9.1430  &  10.3119  \\
\hline
\end{tabular}
\end{center}
\end{table}

\subsection{Fitting the Potentials}

The \ai\ PECs were refined by fitting to the spectroscopic data
identified in Table~\ref{tab:obsdata}. However, extending the temperature
range of the spectra requires consideration of highly excited levels and
extrapolation of the PECs beyond the region determined by
experimental input values, hence care needs to be taken to ensure the
curves maintain physical shapes outside the experimentally refined
regions. In this context we define a physical shape to be the shape of
the \ai\ curve. We tested multiple potential energy forms, namely the
extended Morse oscillator (EMO), Morse long range (MLR) and Morse
Lenard Jones (MLJ) potentials \cite{lr11}, to achieve an optimum fit to the
experimental data whilst maintaining a physical curve shape. Data for
multiple isotopologues were fitted simultaneously to ensure the
resulting curves are valid for all isotopologues. $r_{e}$ and $D_{e}$
were held constant in the fits, as the fits were found to be unstable
otherwise.

For NaCl, BOB terms did not improve the quality of the fit and were
not pursued. Of the 1370 lines used in the fit, 1060 were Na$^{35}$Cl
and 310 were Na$^{37}$Cl. The final potential was expressed as an EMO:
\begin{equation}
 V(r) = D_{e} \left[ 1 - e^{-\phi(r)\left(r-r_{e}\right)} \right]^2,
\end{equation}
where
\begin{equation}
\phi(r) = \sum_{i=0}^N \phi_i y_p(r,r_{e})^i,
\end{equation}
\begin{equation}
 y_p(r,r_{e}) = \frac{r^p - r_{e}^p}{r^p + r_{e}^p}
\end{equation}
and $p$ was set to 3, $N$ to 4, $D_{e}$ to 34120.0 cm$^{-1}$
\citep{HerzHub}, and $r_{e}$ to 2.360796042~\AA\
\citep{97RaDuGu.NaCl}. Parameters resulting from the fit are given in
Table~\ref{tab:potfit1}.

For KCl non-adiabatic BOB terms were included in the fit as they
resulted in an improvement. Of the 549 lines used in the fit, 361 were
$^{39}$K$^{35}$Cl, 82 were $^{39}$K$^{37}$Cl, 64 were
$^{41}$K$^{35}$Cl and 40 were $^{41}$K$^{37}$Cl. The final potential
was expressed as a MLR:
\begin{equation}
V(r) = D_{e} \left[1 - \frac{u_{\rm LR}(r)}{u_{\rm LR}(r_{e})}e^{-\phi(r)\left(r-r_{e}\right)} \right]^{2},
\end{equation}
where
\begin{equation}
\phi(r) = \sum_{i=0}^{N} \phi_{i} y_{p}(r,r_{e})^{i} + y_{p}(r,r_{e})\phi_{\infty},
\end{equation}
\begin{equation}
 y_{p}(r,r_{e}) = \frac{r^{p} - r_{e}^{p}}{r^{p} + r_{e}^{p}}
\end{equation}
\begin{equation}
u_{\rm LR}(r) = \frac{C_{m}}{r^{m}} + \frac{C_{n}}{r^{n}}
\end{equation}
and $p$ was set to 2, $N$ to 3, $m$ to 2, $n$ to 3, $C_{2}$ to 10000, $C_{3}$ to 13000000, $D_{e}$ to 34843.15 \cm\ \citep{61BrBrxx.KCl} and $r_{e}$ to 2.6667253989 \AA\ \citep{04CaLeMu.KCl}.

The non-adiabatic BOB correction function is defined as:

\begin{equation}
g(r) = \frac{M^{ref}}{M} \left[y_{p}(r,r_{e})t_{\infty} + \left[1 - y_{p}(r,r_{e})\right] \sum_{j=0}^{N} t_{j} \left[y_{p}(r,r_{e})\right]^{j} \right]
\end{equation}
where
\begin{equation}
y_{p}(r,r_{e}) = \frac{r^{p} - r_{e}^{p}}{r^{p} + r_{e}^{p}}
\end{equation}
and $M$ is the mass of the particular isotopologue, $M_{ref}$ is the
mass of the parent isotopologue, $p$ was set to 2 and $N$ to 1.
Parameters resulting from this fit are given in
Table~\ref{tab:potfit2}.

\begin{figure}
\begin{center}
\scalebox{0.4}{\includegraphics{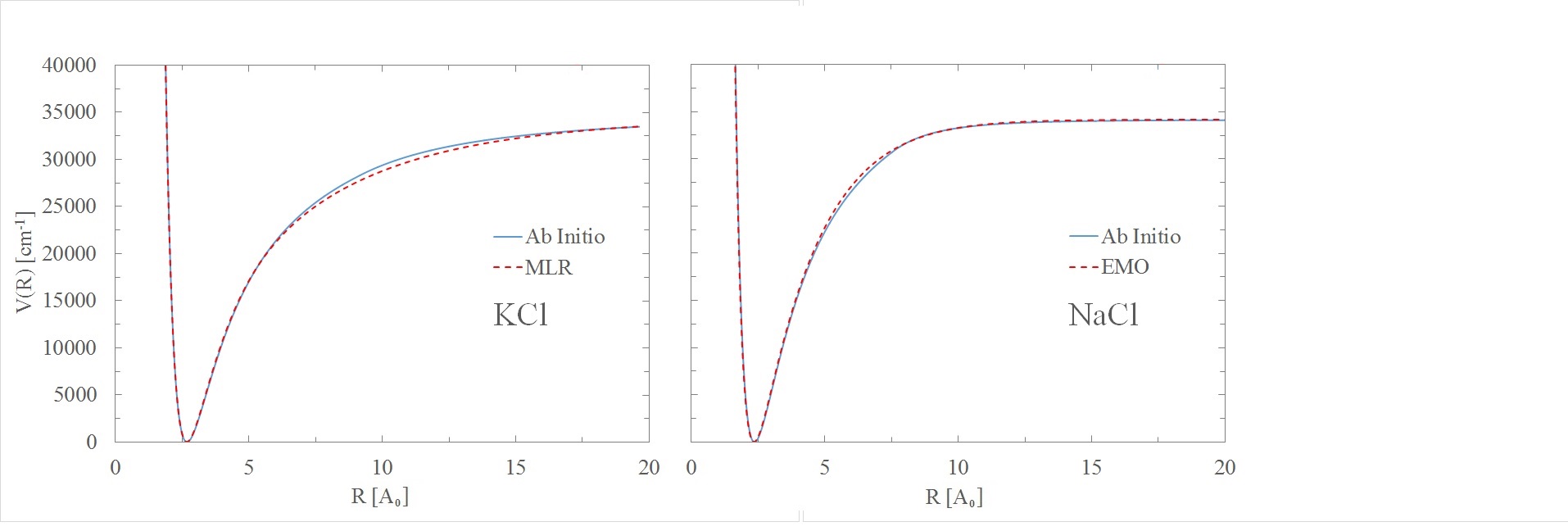}}
\caption{Comparison of \ai\ and fitted ground electronic state potential energy curves for NaCl (left) and KCl (right).}
\label{fig:pot}
\end{center}
\end{figure}

The input experimental data was reproduced within 0.01~\cm\ and often
much better than this. The final curves, shown in Fig~\ref{fig:pot},
follow the \ai\ shape with the exception of
regions 6\AA\ -- 17\AA\ for KCl and 4.5\AA\ -- 8\AA\ for NaCl. These regions
are associated with the textbook avoided crossings
between Columbic X$^+$ -- Cl$^-$ and neutal  X -- Cl PECs which occurs
in the adiabatic
representation of the ground electronic state, see \citet{04GiYo} for a detailed discussion. Without experimental
data near dissociation it is difficult to represent this accurately
with DPOTFIT. Consequently we decided to limit our line lists to
vibrational states lying below 20,000~\cm\ which do not sample these regions.
This has consequences for the temperature range considered. Based on our
partition sum, see Section 2.5, this range is 0 -- 3000 K.

\begin{table}
\caption{Fitting parameters used in the NaCl Extended Morse Oscillator potential,
see eq.~(1). (Uncertainties are given in parenthesis in units of the last
digit.)}
\label{tab:potfit1} \footnotesize
\begin{center}
\begin{tabular}{ll}
\hline
$N$ & $\phi_i$\\
\hline
0 & 0.8947078(17)\\
1 & -0.287528(48)\\
2 & 0.00581(11)\\
3 & -0.0278(14)\\
4 & -0.0290(37)\\
\hline
\end{tabular}
\end{center}
\end{table}

\begin{table}
\caption{Fitting parameters used in the KCl Morse Long Range potential,
see eq.~(4). (Uncertainties are given in parenthesis in units of the last
digit.)}
\label{tab:potfit2} \footnotesize
\begin{center}
\begin{tabular}{lll}
\hline
$N$ & $\phi_{i}$ & $t_{j}$ \\
\hline
0 & -9.075210(10) & 0.0 \\
1 & 1.23590(85)  & 0.00030(12) \\
2 & 0.4859(22) & 0.0 \\
3 & 1.200(10) & 0.0 \\
\hline
\end{tabular}
\end{center}
\end{table}

Comparisons with observed frequencies for Na$^{35}$Cl and
$^{39}$K$^{35}$Cl are given in Table~\ref{tab:energies1} and
\ref{tab:energies2}. These demonstrate the accuracy of fits.
An important aim
in refining a PEC is to also predict spectroscopic data outside the
experimental range. This can be tested for KCl for which there are R
band head measurements up to $v$ = 12 \citep{97RaDuGu.NaCl}. The
positions of these band heads, which are key features in any weak
or low-resolution spectrum, are predicted to high accuracy,
see Table~\ref{tab:bandheads}.
\begin{table}
\caption{Comparison of theoretically predicted Na$^{35}$Cl ro-vibrational wavenumbers, in cm$^{-1}$,
with some of the laboratory measurements of \protect\citet{97RaDuGu.NaCl}. }
\label{tab:energies1} 
\begin{center}
\begin{tabular}{lrlrllr}

\hline

 $v\p$    & $J\p$    &       $v\pp$  &  $J\pp$ &     Obs.   &        Calc.   &  Obs.-Calc.  \\
\hline
1	&	99	&	0	&	98	&	387.0444	&	387.0446	&	-0.0002	\\
1	&	100	&	0	&	99	&	387.1219	&	387.1221	&	-0.0002	\\
1	&	101	&	0	&	100	&	387.1950	&	387.1957	&	-0.0007	\\
2	&	3	&	1	&	2	&	358.9248	&	358.9260	&	-0.0012	\\
2	&	4	&	1	&	3	&	359.3419	&	359.3444	&	-0.0025	\\
2	&	5	&	1	&	4	&	359.7587	&	359.7596	&	-0.0009	\\
3	&	110	&	2	&	109	&	380.2722	&	380.2746	&	-0.0024	\\
3	&	111	&	2	&	110	&	380.3014	&	380.3075	&	-0.0061	\\
3	&	112	&	2	&	111	&	380.3372	&	380.3365	&	0.0007	\\
4	&	28	&	3	&	27	&	361.3425	&	361.3429	&	-0.0004	\\
4	&	29	&	3	&	28	&	361.6718	&	361.6730	&	-0.0012	\\
4	&	31	&	3	&	30	&	362.3244	&	362.3230	&	0.0014	\\
5	&	3	&	4	&	2	&	348.6060	&	348.6104	&	-0.0044	\\
5	&	4	&	4	&	3	&	349.0193	&	349.0195	&	-0.0002	\\
5	&	5	&	4	&	4	&	349.4289	&	349.4254	&	0.0035	\\
6	&	114	&	5	&	113	&	369.5373	&	369.5406	&	-0.0033	\\
6	&	115	&	5	&	114	&	369.5551	&	369.5562	&	-0.0011	\\
6	&	117	&	5	&	116	&	369.5781	&	369.5758	&	0.0023	\\
7	&	73	&	6	&	72	&	362.1649	&	362.1606	&	0.0043	\\
7	&	74	&	6	&	73	&	362.3244	&	362.3276	&	-0.0032	\\
7	&	75	&	6	&	74	&	362.4924	&	362.4910	&	0.0014	\\
8	&	38	&	7	&	37	&	350.7058	&	350.7009	&	0.0049	\\
8	&	39	&	7	&	38	&	350.9881	&	350.9876	&	0.0005	\\
8	&	40	&	7	&	39	&	351.2687	&	351.2710	&	-0.0023	\\
\hline
\end{tabular}
\end{center}
\end{table}

\begin{table}
  \caption{Comparison of theoretically predicted $^{39}$K$^{35}$Cl ro-vibrational wavenumbers, in cm$^{-1}$,
with some of the laboratory data of \protect\citet{97RaDuGu.NaCl}, as re-assigned in this work.}
\label{tab:energies2} 
\begin{center}
\begin{tabular}{lrlrllr}
\hline

  $v\p$    &  $J\p$    &   $v\pp$  &    $J\pp$ &     Obs.   &        Calc.   &  Obs.-Calc.  \\
\hline
1	&	102	&	0	&	101	&	294.9349	&	294.9347	&	0.0002	\\
1	&	103	&	0	&	102	&	295.0173	&	295.0154	&	0.0019	\\
1	&	104	&	0	&	103	&	295.0955	&	295.0941	&	0.0014	\\
1	&	105	&	0	&	104	&	295.1729	&	295.1710	&	0.0019	\\
2	&	43	&	1	&	42	&	284.5787	&	284.5795	&	-0.0008	\\
2	&	49	&	1	&	48	&	285.6588	&	285.6551	&	0.0037	\\
2	&	51	&	1	&	50	&	286.0004	&	286.0001	&	0.0003	\\
2	&	52	&	1	&	51	&	286.1646	&	286.1700	&	-0.0054	\\
3	&	121	&	2	&	120	&	291.1544	&	291.1554	&	-0.0010	\\
3	&	122	&	2	&	121	&	291.2013	&	291.1992	&	0.0021	\\
3	&	123	&	2	&	122	&	291.2433	&	291.2411	&	0.0022	\\
3	&	126	&	2	&	125	&	291.3562	&	291.3555	&	0.0007	\\
4	&	74	&	3	&	73	&	284.6069	&	284.6019	&	0.0050	\\
4	&	75	&	3	&	74	&	284.7309	&	284.7300	&	0.0009	\\
4	&	76	&	3	&	75	&	284.8579	&	284.8563	&	0.0016	\\
4	&	78	&	3	&	77	&	285.1054	&	285.1036	&	0.0018	\\
5	&	111	&	4	&	110	&	285.7192	&	285.7174	&	0.0018	\\
5	&	113	&	4	&	112	&	285.8341	&	285.8375	&	-0.0034	\\
5	&	114	&	4	&	113	&	285.8912	&	285.8947	&	-0.0035	\\
5	&	116	&	4	&	115	&	286.0004	&	286.0037	&	-0.0033	\\
6	&	73	&	5	&	72	&	279.6821	&	279.6817	&	0.0004	\\
6	&	75	&	5	&	74	&	279.9388	&	279.9355	&	0.0033	\\
6	&	76	&	5	&	75	&	280.0552	&	280.0598	&	-0.0047	\\
6	&	81	&	5	&	80	&	280.6586	&	280.6552	&	0.0034	\\
\hline
\end{tabular}
\end{center}
\end{table}

\begin{table}
\caption{Comparison of theoretically predicted $^{39}$K$^{35}$Cl R-branch band heads, in cm$^{-1}$,
with laboratory measurements from \protect\citet{97RaDuGu.NaCl} and this work.}
\label{tab:bandheads}
\begin{center}
\begin{tabular}{clll}
\hline
Band & Observed & Calculated & O -- C \\
\hline
1 -- 0	&	296.702	&	296.703	&	-0.001	\\
2 -- 1	&	294.181	&	294.182	&	-0.001	\\
3 -- 2	&	291.680	&	291.682	&	-0.002	\\
4 -- 3	&	289.201	&	289.203	&	-0.002	\\
5 -- 4	&	286.742	&	286.745	&	-0.003	\\
6 -- 5	&	284.303	&	284.306	&	-0.003	\\
7 -- 6	&	281.884	&	281.887	&	-0.003	\\
8 -- 7	&	279.488	&	279.489	&	-0.001	\\
9 -- 8	&	277.110	&	277.110	&	0.0	\\
10 -- 9	&	274.752	&	274.752	&	0.0	\\
11 -- 10	&	272.414	&	272.411	&	0.003	\\
12 -- 11	&	270.120	&	270.090	&	0.03	\\
\hline
\end{tabular}
\end{center}
\end{table}

\subsection{Partition Functions}

The calculated energy levels, see Section 3, were summed in Excel to
generate partition function values for a range of temperatures.
We determined that our partition function is at least 95$\%$ converged
at 3000 K and much better than this at lower temperatures.
 Therefore temperatures up to 3000~K were considered.
Values for the parent isotopologues are compared to
previous studies, namely \citet{81Irxxxx.partfunc},
\citet{84SaTaxx.partfunc} and CDMS, in Table~\ref{tab:pf}.

For ease of use, we fitted our partition functions, Q, to a series expansion of the form used by \citet{jt263}:
\begin{equation}
\log_{10} Q(T) = \sum_{n=0}^6 a_n \left[\log_{10} T\right]^n \label{eq:pffit}
\end{equation}
with the values given in Table~\ref{tab:pffit}.

\begin{table}
\caption{Comparison of Na$^{35}$Cl and $^{39}$K$^{35}$Cl partition functions}
\label{tab:pf} \footnotesize
\begin{center}
\begin{tabular}{llcrr}
\hline
T(K) & This work& CDMS & \citet{81Irxxxx.partfunc} & \citet{84SaTaxx.partfunc}   \\
\hline
&&&& \\
&& Na$^{35}$Cl && \\
&&&& \\
9.375	&	30.3338	&	30.3307	&	-	&	-	\\
18.75	&	60.3352	&	60.3299	&	-	&	-	\\
37.5	&	120.3556	&	120.3455	&	-	&	-	\\
75	&	240.6984	&	240.6770	&	-	&	-	\\
150	&	496.6455	&	496.5538	&	-	&	-	\\
225	&	802.3712	&	802.1167	&	-	&	-	\\
300	&	1173.0397	&	1172.5403	&	-	&	-	\\
500	&	2506.9232	&	2505.0340	&	-	&	-	\\
1000	&	8161.702	&	-	&	8204.6	&	8165.4	\\
1500	&	17333.48	&	-	&	17409.8	&	16960.3	\\
2000	&	30294.77	&	-	&	30370.1	&	29685.3	\\
2500	&	47362.31	&	-	&	47324.9	&	46807.2	\\
3000	&	68909.60	&	-	&	68530.1	&	68766.1	\\
&&&& \\
&& $^{39}$K$^{35}$Cl && \\
&&&& \\
9.375	&	51.1529	&	51.1495	&	-	&	-	\\
18.75	&	101.9823	&	101.9724	&	-	&	-	\\
37.5	&	203.6737	&	203.6504	&	-	&	-	\\
75	&	409.1563	&	409.1053	&	-	&	-	\\
150	&	876.2078	&	876.0902	&	-	&	-	\\
225	&	1474.9611	&	1474.7618	&	-	&	-	\\
300	&	2225.1732	&	2224.8905	&	-	&	-	\\
500	&	5000.7402	&	5000.3352	&	-	&	-	\\
1000	&	17092.48	&	-	&	17277.73	&	17112.5	\\
1500	&	36808.46	&	-	&	37327.7	&	36147.1	\\
2000	&	64110.97	&	-	&	65747.4	&	64142.9	\\
2500	&	99023.43	&	-	&	103058.6	&	102212.0	\\
3000	&	142068.17	&	-	&	149837.7	&	151368.0	\\
\hline
\end{tabular}
\end{center}
\end{table}

\begin{table}
\caption{Fitting parameters used to fit the partition functions,
see eq.~\ref{eq:pffit}. Fits are valid for temperatures between 500 and 3000 K.}
\label{tab:pffit} \footnotesize
\begin{center}
\begin{tabular}{lrrrrrr}
\hline
&  Na$^{35}$Cl & Na$^{37}$Cl &  $^{39}$K$^{35}$Cl   &  $^{39}$K$^{37}$Cl    &  {$^{41}$K$^{35}$Cl} & $^{41}$K$^{37}$Cl \\
\hline
$a_0$	&	35.528812	&	39.941335	&	372.555818	&	72.206029	&	74.926932	&	72.591531	\\
$a_1$	&	-65.142353	&	-73.36368	&	-762.782356	&	-138.3407430	&	-143.689312	&	-139.093018	\\
$a_2$	&	53.290409	&	59.6576584	&	653.21703	&	114.10145500	&	118.474798	&	114.705511	\\
$a_3$	&	-23.592248	&	-26.212185	&	-297.787108	&	-50.41476400	&	-52.319838	&	-50.665015	\\
$a_4$	&	6.036705	&	6.64133762	&	76.294719	&	12.68394200	&	13.1500522	&	12.740842	\\
$a_5$	&	-0.8370958	&	-0.9113382	&	-10.405679	&	-1.71634070	&	-1.7770609	&	-1.7230912	\\
$a_6$	&	0.04887272	&	0.05266306	&	0.5898907	&	0.097426548	&	0.1007164997	&	0.0977526634	\\
\hline
\end{tabular}
\end{center}
\end{table}

\subsection{Line-List Calculations}

While sodium has only a single stable isotope, $^{23}$Na,
both potassium and chlorine each have two:
$^{39}$K (whose natural terrestrial abundance is about 93.25\%) and
 $^{41}$K (6.73\%), and $^{35}$Cl (75.76\%) and $^{37}$Cl (24.24\%).
Line lists were therefore calculated for two NaCl and four KCl isotopologues.
Ro-vibrational states up to $v$ = 100, $J$ = 563 and $v$ = 120, $J$ =
500 respectively, and all transitions between these states satisfying
the dipole selection rule $\Delta J = \pm 1$, were considered. A
summary of each line list is given in Table~\ref{tab:finallevels}.

The procedure described above was used to produce line lists, i.e.
catalogues of transition frequencies $\nu_{ij}$ and Einstein
coefficients $A_{ij}$ for two NaCl and four KCl isotopologues:
Na$^{35}$Cl, Na$^{37}$Cl, $^{39}$K$^{35}$Cl, $^{39}$K$^{37}$Cl,
$^{41}$K$^{35}$Cl and $^{41}$K$^{37}$Cl. The computed line lists are
available in electronic form as supplementary information to this
article.

\begin{table}
\caption{Summary of our line lists.}
\label{tab:finallevels}
\begin{center}
\begin{tabular}{lrrrrrr}
\hline
& Na$^{35}$Cl & Na$^{37}$Cl &  $^{39}$K$^{35}$Cl   &  $^{39}$K$^{37}$Cl    &  {$^{41}$K$^{35}$Cl} & $^{41}$K$^{37}$Cl \\
\hline
Maximum $v$ & 100 & 100 & 120 & 120 & 120 & 120 \\
Maximum $J$ & 557 & 563 & 500 & 500 & 500 & 500 \\
Number of lines& 4734567& 4763324 & 7224331 & 7224331 & 7224331 & 7224331  \\
\hline
\end{tabular}
\end{center}
\end{table}

\section{Results}
The full line list computed for all isotopologue considered is
summarised in Table~\ref{tab:finallevels}.  Each line list contains around 4
-- 7 million transitions and are therefore, for compactness and
ease of use, divided into separate energy files and transition files.
This is done using the standard ExoMol format \citep{jt548} which is
based on a method originally developed for the BT2 line list
\citep{jt378}. Extracts from the start of the Na$^{35}$Cl and
$^{39}$K$^{35}$Cl files are given in Table~\ref{tab:states1},
Table~\ref{tab:trans1}, Table~\ref{tab:states2} and
Table~\ref{tab:trans2}. They can be downloaded from the CDS via XXX.
The line lists and partition functions can also be obtained from
www.exomol.com.

\begin{table}
\caption{Extract from start of states file for Na$^{35}$Cl}
\label{tab:states1}
\begin{center}
\begin{tabular}{llrll}
\hline
$I$ &   $\tilde{E}$      &  $g$  &  $J$ & $v$\\
\hline
           1   &  0.000000    & 16  &     0  &  0 \\
           2    & 0.434501  &   48  &     1   & 0 \\
           3    & 1.303497 &    80   &    2  &  0 \\
           4    & 2.606971  &  112   &    3   & 0 \\
           5    & 4.344901   & 144   &    4   & 0 \\
           6    & 6.517259 &  176   &    5 &   0 \\
\hline
\end{tabular}

\noindent
  $I$:   State counting number;
 $\tilde{E}$: State energy in \cm;
  $g$: State degeneracy;
$J$:   State rotational quantum number;
 $v$:   State vibrational quantum number.

\end{center}
\end{table}

\begin{table}
\caption{Extract from start of states file for $^{39}$K$^{35}$Cl}
\label{tab:states2}
\begin{center}
\begin{tabular}{llrll}
\hline
$I$ &   $\tilde{E}$      &  $g$  &  $J$ & $v$\\
\hline
1	&	0.0	        &	16	&	0	&	0	\\
2	&	0.256466	&	48	&	1	&	0	\\
3	&	0.769393	&	80	&	2	&	0	\\
4	&	1.538778	&	112	&	3	&	0	\\
5	&	2.564613	&	144	&	4	&	0	\\
6	&	3.846887	&	176	&	5	&	0	\\
\hline
\end{tabular}

\noindent
  $I$:   State counting number;
 $\tilde{E}$: State energy in \cm;
  $g$: State degeneracy;
$J$:   State rotational quantum number;
 $v$:   State vibrational quantum number.

\end{center}
\end{table}

\begin{table}
\caption{Extracts from the transitions file for Na$^{35}$Cl}
\label{tab:trans1}
\begin{center}
\begin{tabular}{lll}
\hline
       $I$  &  $F$ & $A_{IF}$\\
\hline
2	&	1	&	7.21E-07	\\
3	&	2	&	6.93E-06	\\
4	&	3	&	2.50E-05	\\
5	&	4	&	6.16E-05	\\
6	&	5	&	1.23E-04	\\
7	&	6	&	2.16E-04	\\
\hline
\end{tabular}

\noindent
 $I$: Upper state counting number;
$F$:      Lower state counting number;
$A_{IF}$:  Einstein A coefficient in s$^{-1}$.

\end{center}
\end{table}

\begin{table}
\caption{Extracts from the transitions file for $^{39}$K$^{35}$Cl}
\label{tab:trans2}
\begin{center}
\begin{tabular}{lll}
\hline
       $I$  &  $F$ & $A_{IF}$\\
\hline
2	&	1	&	1.89E-07	\\
3	&	2	&	1.81E-06	\\
4	&	3	&	6.55E-06	\\
5	&	4	&	1.61E-05	\\
6	&	5	&	3.21E-05	\\
7	&	6	&	5.64E-05	\\
\hline
\end{tabular}

\noindent
 $I$: Upper state counting number;
$F$:      Lower state counting number;
$A_{IF}$:  Einstein A coefficient in s$^{-1}$.

\end{center}
\end{table}

\begin{figure}
\begin{center}
\scalebox{0.4}{\includegraphics{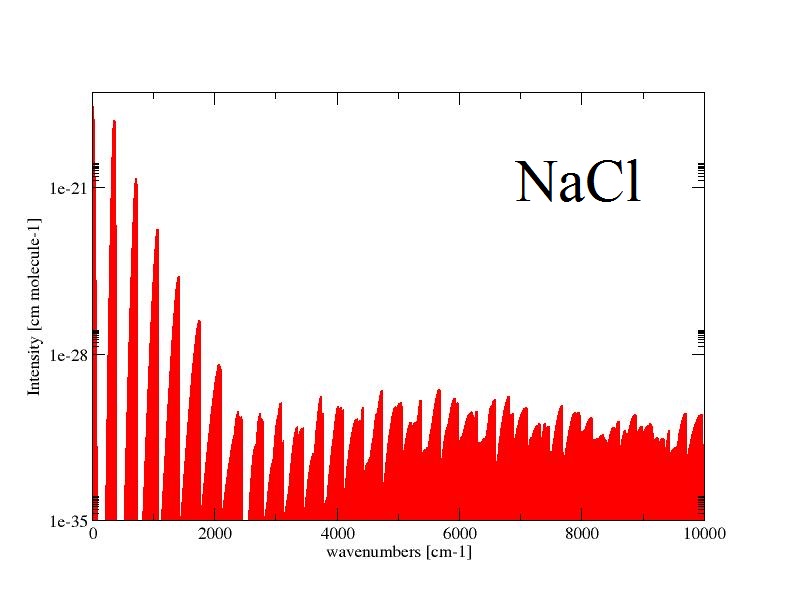}}
\scalebox{0.4}{\includegraphics{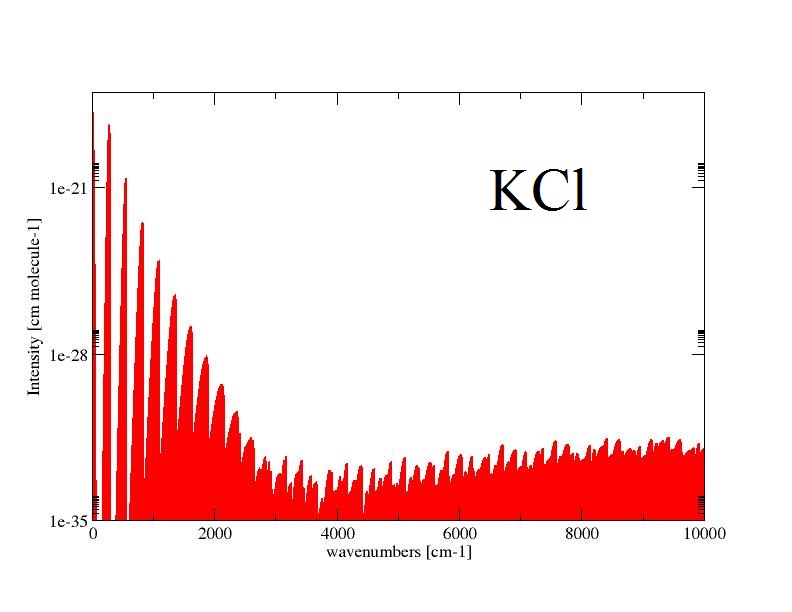}}
\caption{Absorption spectra of Na$^{35}$Cl and $^{39}$K$^{35}$Cl at 300 K.}
\label{fig:linelist}
\end{center}
\end{figure}

Figure~\ref{fig:linelist} illustrates
the synthetic absorption spectra of Na$^{35}$Cl and $^{39}$K$^{35}$Cl
at 300 K.  As
the dipole moment curves are essentially straight lines, the overtone bands
for these molecules are very weak meaning that key spectral features
are confined to long wavelengths associated with the pure
rotational spectrum and the vibrational fundamental.

\begin{figure}
\begin{center}
\scalebox{0.4}{\includegraphics{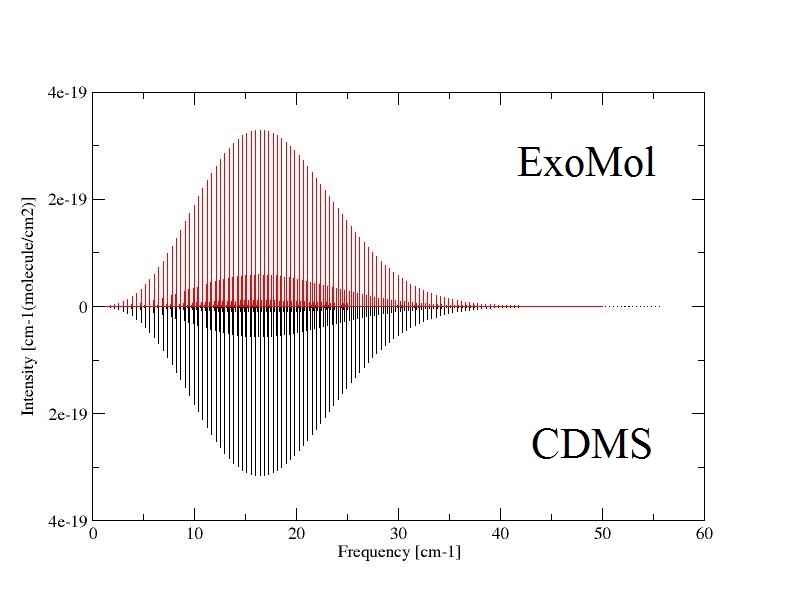}}
\scalebox{0.4}{\includegraphics{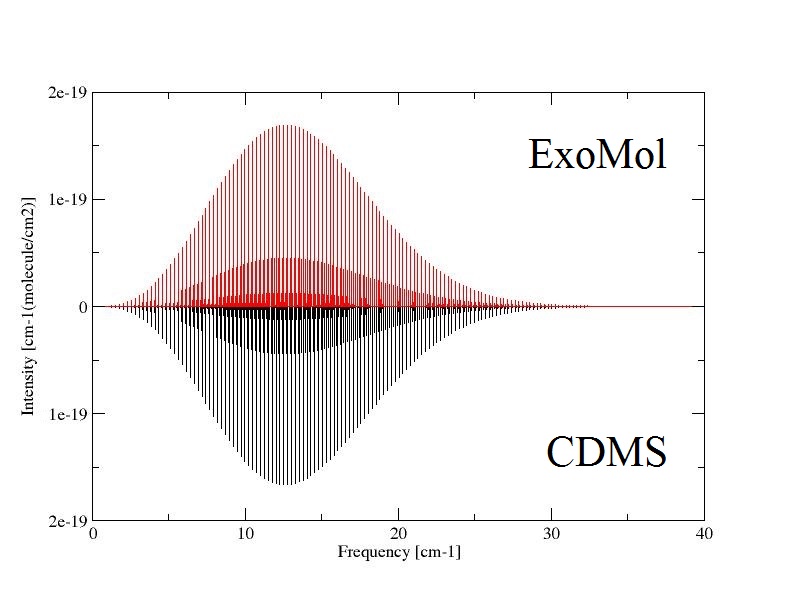}}
\caption{Absorption lines of Na$^{35}$Cl and $^{39}$K$^{35}$Cl at 300 K: ExoMol vs. CDMS.}
\label{fig:CDMScomp}
\end{center}
\end{figure}

The CDMS database contains 607 and 772 rotational lines for
Na$^{35}$Cl and $^{39}$K$^{35}$Cl respectively.  Comparisons with the
CDMS lines are presented in Figure~\ref{fig:CDMScomp}. As can be seen the agreement is excellent
for both frequency and intensity. In particular,
 predicted  line intensities agree within
2$\%$ and 4$\%$ for the KCl and NaCl isotopomers considered in CDMS respectively.

\begin{figure}
\begin{center}
\scalebox{0.4}{\includegraphics{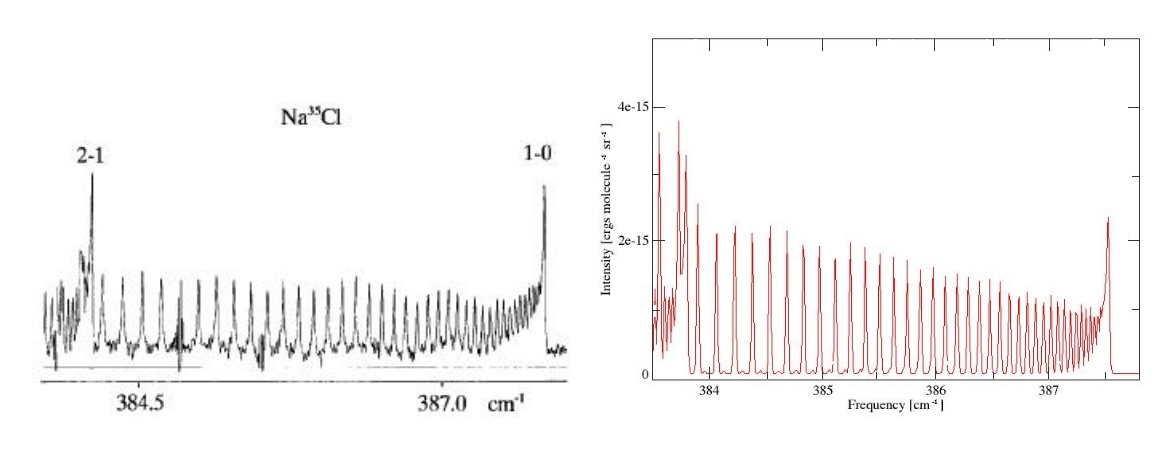}}
\caption{Emission spectra of NaCl at 1273 K: left, \protect\citet{97RaDuGu.NaCl}; right, ExoMol. (Reprinted from Ref. \citep{97RaDuGu.NaCl}. Copyright 1997, with permission from Elsevier.) }
\label{fig:exp1}
\end{center}
\end{figure}

\begin{figure}
\begin{center}
\scalebox{0.4}{\includegraphics{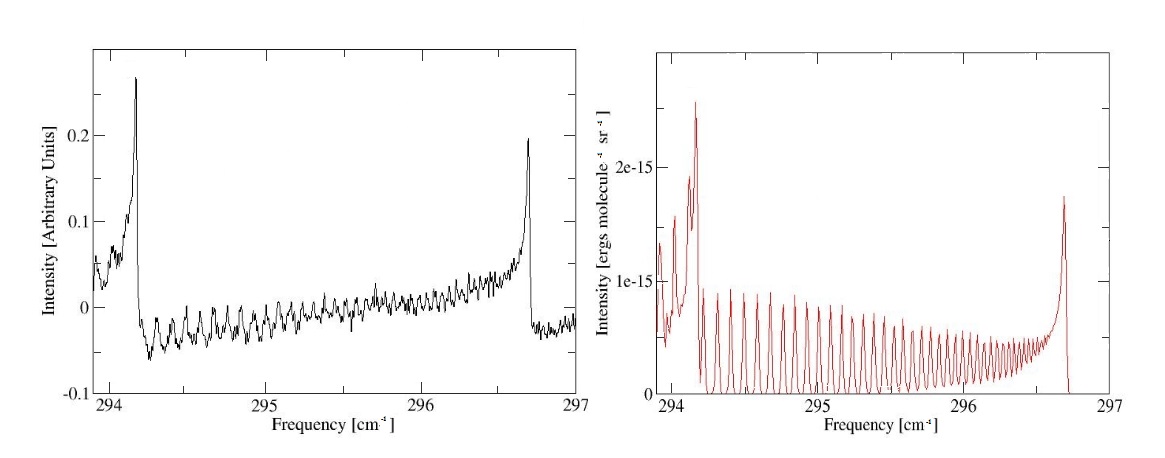}}
\caption{Emission spectra of KCl at 1273 K:left, \protect\citet{97RaDuGu.NaCl}; right, ExoMol.}
\label{fig:exp2}
\end{center}
\end{figure}

Emission cross-sections for Na$^{35}$Cl and $^{39}$K$^{35}$Cl
were simulated using Gaussian line-shape profiles with half-width = 0.01
cm$^{-1}$ as described by \citet{jt542}. The resulting synthetic emission
spectra are compared to the experimental ones in Figs.~\ref{fig:exp1}
and \ref{fig:exp2}. When making comparisons one has to be aware of a
number of experimental issues.  The baseline in NaCl shows residual
"channeling": a sine-like baseline that often appears in FTS spectra
due to interference from reflections from parallel optical surfaces in
the beam.  For KCl the spectrum is very weak and the baseline, which
has a large offset, was not properly adjusted to zero. Given these
considerations, the comparisons must be regarded as satisfactory.

\section{conclusions}

We present accurate but
comprehensive line lists for the stable isotopologues of
NaCl and KCl. Laboratory frequencies are
reproduced to much more than sub-wavenumber accuracy.
This accuracy should extend to
all predicted transition frequencies up to at least $v$ = 8 and $v$ =
12 for NaCl and KCl respectively. New \ai\ dipole moments and Einstein
$A$ coefficients are computed. Comparisons with the semi-empirical CDMS
database suggest the intensities the pure rotational  are accurate.

The results are line lists for the rotation-vibration transitions
within the ground states of Na$^{35}$Cl, Na$^{37}$Cl,
$^{39}$K$^{35}$Cl, $^{39}$K$^{37}$Cl, $^{41}$K$^{35}$Cl and
$^{41}$K$^{37}$Cl, which should be accurate for a range of
temperatures up to at least 3000~K. The line lists can be downloaded
from the CDS or from www.exomol.com.

Finally, we note that HCl is likely to be the other main
chlorine-bearing species in exoplanets.  A comprehensive line lists
for H$^{35}$Cl and H$^{37}$Cl have recently been provided by
\citet{13LiGoHa.HCl} and \citet{13LiGoLe.HCl}.

\section*{Acknowledgements}

We thank Alexander Fateev for stimulating discussions and Kevin
Kindley for some preliminary work with the KCl infrared emission
spectrum.  This work was supported by grant from energinet.dk under a
subcontract from the Danish Technical University and by the ERC under
the Advanced Investigator Project 267219.
Support was also provided by the NASA Origins of Solar Systems program.


\label{lastpage}

\end{document}